# Current-Driven Magnetic Memory with Tunable Magnetization Switching

S.K. Wong, A.B. Pakhomov, S.T. Hung, S.G. Yang, and C.Y. Wong

Magnetics Innovation Center (MAGIC), Materials Characterization and Preparation Facility, Hong Kong University of Science and Technology, Clear Water Bay, Kowloon, Hong Kong

Co(x nm, x=10nm or 40nm)/Cu(5nm)/Co(2.5nm) layers were deposited between copper electrodes in $SiO_2$ vias. Magnetic states, and the corresponding resistance states, of these devices were switched by electric currents perpendicular to the layers. The I-V loops show asymmetric behavior with hysteresis. When electrons flow in the direction from thick to thin Co layer (positive current), multiple switches were observed on increasing current up to a chosen maximum positive $I_{write}$. On decreasing current from $I_{write}$, the I-V curve was smooth and characterized by considerably lower resistance. Under reverse current, an abrupt switch to the high resistance state occurred at the current value $I_{erase} \approx -0.9 \times I_{write}$. Resistance had a maximum at zero current in both states, where the ratio $R_{high}/R_{low}$ could be as high as factor of four.

Several designs of magnetic random access memory based on either giant magnetoresistance (GMR) or magnetic tunnel junctions (MTJ) elementary cells have been proposed recently (for example, Refs. 1 and 2). In these proposals of magnetic memories, magnetization reversal of one of the magnetic layers and the related change of resistance (writing) could be achieved by local application of magnetic field of appropriate amplitude and direction. However one of the problems with applied magnetic field is that in a compact array of cells, it is difficult to singly affect the magnetic state of one cell without affecting magnetic states of the other cells. It is advantageous therefore to be able to controllably switch the state of the element by electric current through the element rather than by magnetic field. This principle can be realized in magnetic devices where the magnetization state is changed by high density spin-polarized current flowing perpendicular to the magnetic layers (the effect referred to as spin transfer [3]). In another design the Oersted field associated with current can switch vortex magnetic states in current-perpendicular-to-plane (CPP) GMR geometry. Possibility of switching of magnetization by spin transfer was confirmed in manganite nanometer-scale particles in manganite trilayer junctions [4], in continuous magnetic Co/Cu multilayers, with current injected through a point contact [5], in Co/Cu/Co pillars of diameter about 130±30 nm [6], and in Co nanomagnets in proximity to continuous ferromagnet [7]. Re-alignment of magnetization by the self field of the current was reported in CPP GMR multilayers with diameter 250-500 nm [8, 9]. A small lateral size is important for application of perpendicular memory devices, as it increases the device resistance, decreases the necessary current and potentially allows higher memory array density.

In this paper we demonstrate the feasibility of tunable magnetization switching with vertical Co/Cu/Co current-driven memory devices of lateral dimensions from 50nm to 500nm. The method involves deposition of the desired multilayer directly into vias in an insulator. The I-V loops show asymmetric switching of the resistive state of the device. We find several new features in the I-V behavior compared to the previous works on Co/Cu structures [5-9]. The ratio of resistances in two magnetic states at small sense current can be as large as a factor of four and increases with increasing positive write current $I_{write}$, which can be set at any value up to 100 mA, but which must exceed a small critical value $I_{c+} \approx 2.6$ mA. The negative critical, or erase, current $I_{erase}$ is a dependent parameter, which scales linearly with $I_{write}$ and is thus also well defined.

Bottom Pt contact electrodes were fabricated on an oxidized silicon substrate. $SiO_2$ layer was deposited over the center of the bottom electrode. Vias in $SiO_2$ of dimensions 50×50, 100×100, 250×250, and 500×500 nm$^2$ were etched with Focus Ion Beams. Then Co(x, x=10nm or 40nm)/Cu(5nm )/Co(2.5nm) submicron pillars were deposited into these via openings by e-beam evaporation. Finally a Cu top electrode was deposited perpendicular to the bottom electrode. The idealized cartoon of the device is shown in Fig. 1 (a). Figure 1 (b) shows the configuration of crossing electrodes; the large transparent square is $SiO_2$ layer. Figure 1 (c) shows an example of the rounded 50×50 nm$^2$ via opening. For larger dimensions, the shape was close to a square.

The I-V characteristics of the samples were studied on an HP 4156 B Semiconductor Parameter Analyzer, in the current sweep regime. We define the bottom electrode as ground (Fig. 1 (a)), so that electrons flow from thick to thin Cobalt layer at positive bias. We measured the I-V loops in both symmetric and asymmetric sweeps, namely the maximum positive current was less than or equal to the absolute value of the maximum negative current. Typical sweeps started at a negative current, then the current was increased to a chosen positive value, and then swept back to the starting point. In measurements at current below ~5 mA, about 85% of samples of all dimensions had nearly linear I-V characteristics with resistances distributed between 1 to 50 $\Omega$. This is consistent with the measured value of resistivity of Cu/Co/Cu/Co/Cu films $\rho_{film} \approx 16$ $\mu\Omega$–cm. The remaining 15% of samples showed non-linear behavior, with the maximum resistance of several hundred Ohms at zero current. After application of a higher current, usually in the range ~ 20-100 mA, depending on the sample cross-section and initial resistance, the I-V curves irreversibly turned non-linear in most of the samples.

Figure 2 (a) shows a typical set of three I-V loops, all of which start and complete at –90 mA. In this case, the measurements were done on a sample with a 50 nm via opening. The positive return points were at 90 mA , 40 mA and 10 mA. Arrows show the direction of sweep for the loop of the largest current amplitude. The corresponding resistance R=V/I loops are shown in Fig. 2 (b). Figure 2 (c) shows the differential resistance dV/dI. Imposed on the generally non-linear behavior, the following switching features are observed. All three dependences follow a single curve from –90 mA to a small positive value, which can be defined as the positive critical current $I_{c+}$. At this current the first small discontinuities (downward steps of voltage) can be seen. These discontinuities are best seen as peak features of dV/dI. The value of $I_{c+}$ is a random number unique to each sweep. The smallest value found from many experiments is about 2.6 mA. These partial switches appear at random on increasing current, even up to the largest positive currents available in our setup (100 mA). After the positive current has reached the chosen maximum value (90, 40, or 10 mA), the direction of sweep is reversed. If the maximum current exceeds $I_{c+}$, voltage will now follow a different curve, characterized by smaller values of both V/I and dV/dI. The maximum value of the resistance reached at zero current decreases with the maximum positive current in the loop. The system remains in its low resistance state until a negative switching current is reached, unique for each of the curves, where the system switches back to the high resistive state. It remains in this state indefinitely, unless the current is again increased over the positive value $I_{c+}$. We define the two limiting currents as $I_{write} > I_{c+}$ (positive) and $I_{erase}$ (negative).

Figure 3 demonstrates the write, read and erase processes. The writing current is chosen at $I_{write}$= 50 mA. In writing, current is swept to +50 mA. Reading the low resistance is demonstrated by 100 cycles between –25 and +25 mA. No drift of the resistance values is observed. Erasing (switching to the high resistance state) occurs on sweeping the current to –50 mA, and is seen as a peak in dV/dI. Finally, 7 cycles between –25 and 0 mA imitate the read process in the high-resistance state. Negative current does not effect the high resistance state, and the curves coincide.

The value of the erase current /$I_{erase}$/ is always smaller than $I_{write}$, and is linked to the latter by a simple relationship $I_{erase} \approx -0.9 \times I_{write}$. This is illustrated in Fig. 4. Each point here is the statistical average based on 20 measurements of the positions of the peaks in dV/VI for each value of the write current. In fact, the standard deviation shown by error bars is smaller than the peak width (which is about 2 mA independent on the $I_{write}$). The straight line in Fig. 4 is the best linear fit. The accuracy of the write current in our design is absolute, as it is set at will and can be chosen at any appropriate value. As we see from Fig. 4, $I_{erase}$ is also very well defined by this choice. The high resistance-state curves for all values of $I_{write}$ coincide at any current below $I_{c+}$, including zero, and symmetric at small currents (Fig. 2). On the other hand, the low-resistance state is symmetric with respect to larger currents, but is strongly dependent on $I_{write}$. The ratio of the resistances measured at small currents, $R_{high}/R_{low}$, is surprisingly large. In this sample, Co(x=10nm)/Cu/Co, $R_{high}/R_{low} \approx 4$ when $I_{write}$=90 mA, extrapolating to unity at zero switching field (Fig. 5). Based on statistics collected on 20 samples of different dimensions with Co(x=40nm)/Cu/Co, the values of $R_{high}$ are distributed between 100 and 330 $\Omega$ and are on the average decreasing with increasing the sample size (Fig. 6 (a)).. The ratios $R_{high}/R_{low}$ are in the range between 1.4 and 6.2 (Fig. 6 (b)).

It is interesting to note that one of our samples has been subjected to at least 1000 experimental current cycles with no effect on the performance. One can estimate the following experimental conditions for a 50 nm sample, when the current through the device is I~100 mA ( the maximum value we have used): Current density j ~ $6.4 \times 10^8$ A/cm$^2$; Magnetic field near the channel walls: H ~ $4 \times 10^6$ A/m ~ $3.6 \times 10^4$ Oe; magnetic pressure: P ~ $7.3 \times 10^6$ kG/m$^2$ ~ 7.3 kG/mm$^2$. We also checked that the resistive state of the device is practically not influenced by an external magnetic field up to 10000 Oe. The change of the erase current $I_{erase}$ in this field is comparable to the statistical deviations.

In this paper, we describe the switching events between high and low resistance states as due to changes in magnetization states of the system under the effect of electric current. We notice several discrete steps in voltage with increasing positive current (Fig. 2), which means that several domains participate in the process, while a single abrupt step is observed on switching from low to high resistance state. It is interesting that the multi-domain process is present even when the sample diameter is as small as 50 nm.

To conclude, we have demonstrated a nanometer scale current-controlled magnetic memory device, prepared by deposition of a Co/Cu/Co structure into an SiO$_2$ via. This memory device provides tunable switching between the high resistance and low resistance states, with the flexibility of optional write and erase currents. Additionally, the ratio of resistances in the high and low resistive states can be as high as 4.

This work was supported by the Hong Kong Innovation and Technology Commission, ITF Grant number AF/155/99. The authors would like to thank Micrion for their assistance in the FIB process. We would also like to thank Dr. Stuart Parkin, IBM, Almaden Research Center, for stimulating discussions.

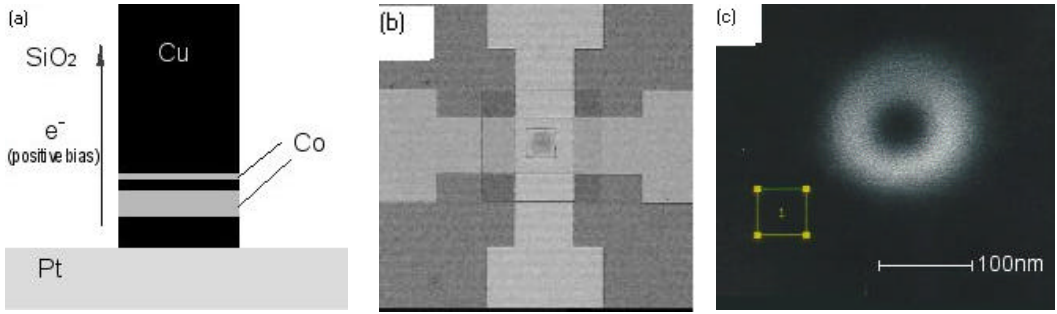

Fig. 1. (a) Schematic cross section of the device; the positive current corresponds to electron flow from bottom to top of the pillar. (b) Crossing electrodes; the large transparent square is $SiO_2$ layer. (c) surface of Silica with a 50 nm opening.

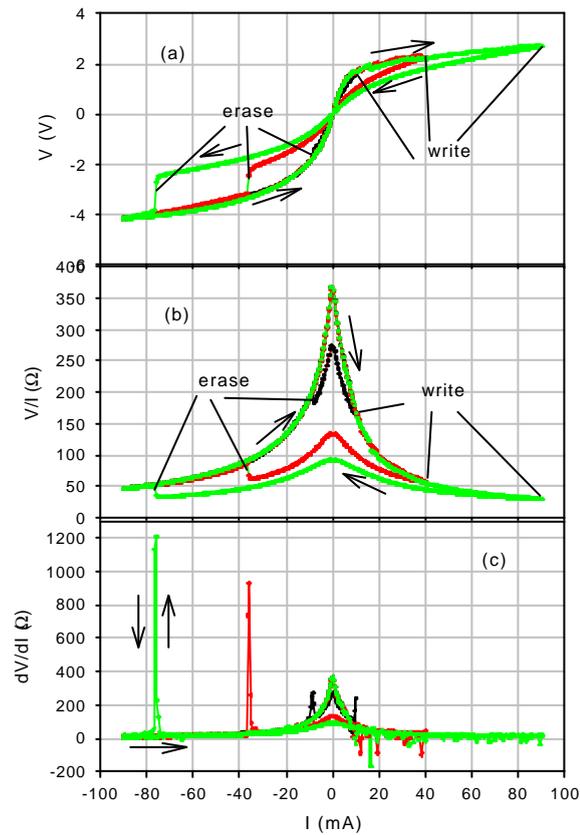

Fig. 2. The voltage (a), resistance (b), and differential resistance (c) as a function of applied current, for three choices of the write current: 90, 40, and 10 mA.

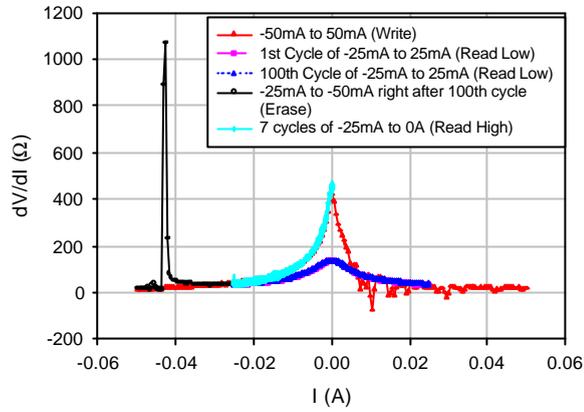

Fig. 3. The write - read - erase cycle.

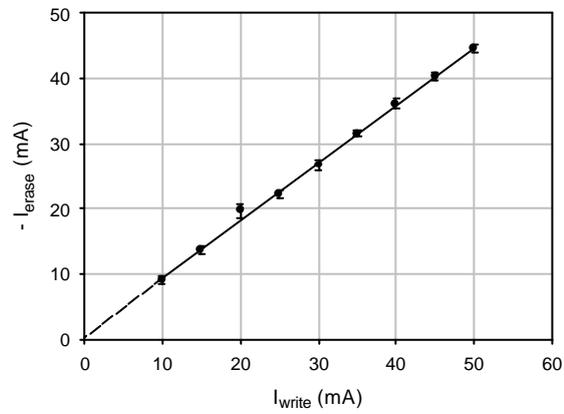

Fig. 4. The (negative) erase current as a function of the chosen write current value. Each point represents 20 measurements of I-V loops. The best fit $I_{erase} \approx -0.88 \times I_{write}$ is shown by the straight line.

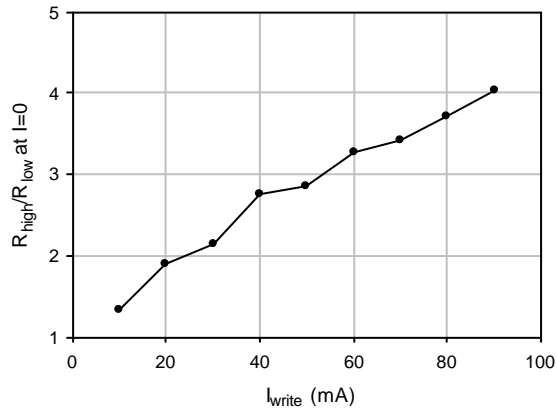

Fig. 5. The ratio of resistances in the two states of the device as a function of the write current.

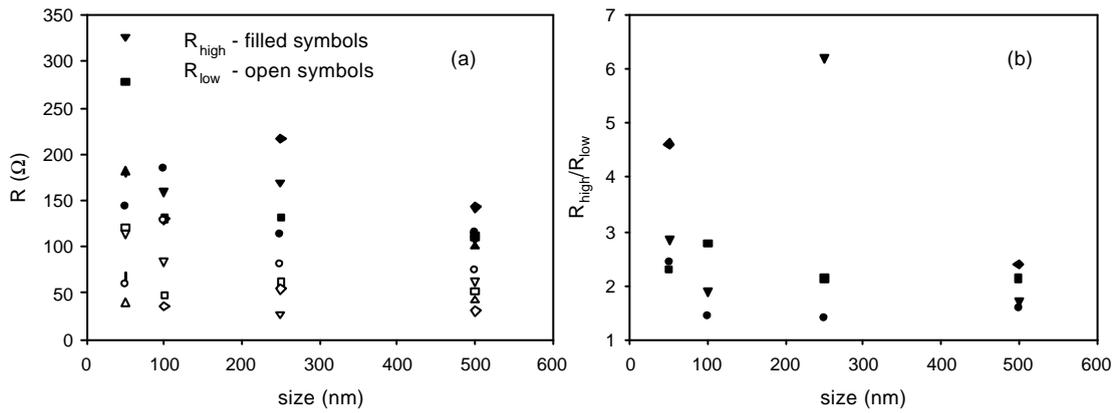

Fig. 6. (a) High resistance (filled symbols) and low resistance (open symbols) of the samples plotted vs the lateral size of the device; (b) the ratio of the high resistance to the low resistance, for the series of Co(x=40nm)/Cu/Co devices.